\documentclass[aps,twocolumn,epsfig,graphics,showpacs,floatfix]{revtex4}

\usepackage{amsmath,amsfonts,amssymb,graphics,graphicx,epsfig,color,times,bbm}

\begin{document}
\bibliographystyle{apsrev}

\newcommand{\proofend}{\hfill\fbox\\\medskip }

\newcommand{\real}{{\text{Re}}}
\newcommand{\imag}{{\text{Im}}}

\newtheorem{theorem}{Theorem}
\newtheorem{proposition}{Proposition}

\newtheorem{lemma}{Lemma}

\newtheorem{definition}{Definition}
\newtheorem{corollary}{Corollary}

\newcommand{\proof}[1]{{\bf Proof.} #1 $\proofend$}

\title{Distillation of continuous-variable entanglement with optical means}

\author{J.\ Eisert$^{1,2}$, D.E.\ Browne$^{2}$, S.\ Scheel$^{2}$, and M.B.\ Plenio$^{2}$}
\affiliation{Institut f{\"u}r Physik, Universit{\"a}t Potsdam,
Am Neuen Palais 10, D-14469 Potsdam, Germany\\
QOLS, Blackett Laboratory, Imperial College London,
Prince Consort Road, London SW7 2BW, UK}

\date{\today}

\begin{abstract}
We present an event-ready
procedure that is capable of distilling Gaussian
two-mode entangled states from a supply of weakly entangled states that have
become mixed in a decoherence process. This procedure relies on
passive optical elements and photon detectors distinguishing the
presence and the absence of photons, but does not make use of
photon counters. We identify fixed points of the
iteration map, and discuss in detail its convergence properties. 
Necessary and sufficient criteria for the convergence to two-mode Gaussian
states are presented.
On the basis of various examples we discuss the performance of
the procedure as far as the increase of the degree of entanglement and
two-mode squeezing is concerned. Finally, we consider imperfect
operations and outline the robustness of the scheme
under non-unit detection efficiencies of the detectors.
This analysis implies that the proposed protocol can be implemented
with currently available technology and detector efficiencies.

\end{abstract}

\draft

\maketitle


\section{Introduction}

The key requirement in essentially all practical  applications of
quantum information science is to protect the coherence of quantum
states against decoherence induced by the uncontrolled influences
of an environment. Entangled states of composite quantum systems,
in particular, deteriorate typically into mixed quantum states
with mere classical correlations. Decoherence accompanies to some
extent any attempt to distribute entangled states that have been
prepared using some local interaction. The first obvious strategy
to minimize this effect is to try to reduce the interaction with
the environment, e.g., by using glass fibers with a long characteristic absorption  
length when distributing photonic entanglement.

This strategy is, however, not enough in many cases. If one
intends to share entanglement over arbitrary distances,
entanglement must in some way be extracted out of the only weakly
entangled quantum systems. Under the key word entanglement
distillation such strategies have been developed \cite{Bennett},
most notably the iterative distillation schemes that work for
spin-$1/2$ or qubit systems. In such schemes, entanglement is
distilled out of a supply of weakly entangled pairs of quantum
systems by means of local operations and classical communication.
Such protocols have been proposed as theoretical concepts
\cite{Bennett}, and discussed under the assumption of imperfect
devices \cite{HansHans}. They form the basis of the theoretical
concept of the distillable entanglement \cite{Rains}, which
clarifies the notion of entanglement as a resource. All the
proposed protocols share nevertheless one property: they are
experimentally very difficult to implement. First steps
towards the full experimental realization of a distillation scheme
very recently been taken \cite{Exp}, based on earlier theoretical
work on optical distillation protocols in the finite-dimensional
setting \cite{Fin}. However, it seems fair to say that the
practical realization of full-scale entanglement distillation
remains one of the key challenges of the field.

In this paper we present a feasible alternative to entanglement
distillation in the finite-dimensional setting. We present a
procedure that is capable of entanglement distillation and
purification employing systems with canonical coordinates, or
so-called continuous-variable systems such as light modes or
harmonic oscillator systems. The scheme that we present is as such
an extension of the Gaussification scheme of Ref.\
\cite{Gaussify}. In this paper, we investigate the power of this
procedure to distill states that have become {\it mixed} as a
result of a decoherence process \cite{Scheel}. More specifically,
we investigate whether Gaussian states that originate for example
from two-mode squeezed states \cite{TMSSTheory,TMSSTheory2,
tmss,GExp} that underwent a decoherence process can be transformed
back into highly entangled Gaussians using feasible local
operations and classical communication only.

The result that, highly entangled and often even pure Gaussian
states can be extracted from a supply of weakly entangled mixed
states, appears to be in strong contrast to the no-go-theorems
concerning the distillation of Gaussian states with Gaussian
operations \cite{NoGo,NoGo2,NoGo3}. However, the central idea is
to leave the Gaussian setting in a first preparation step, and
only in the course of the iteration the resulting states converge
towards a final Gaussian state. In this sense one can sneak
through a loophole left open by the statements demonstrating the
impossibility of entanglement distillation when staying entirely
within the Gaussian setting. It has been shown that the general use of
non-Gaussian operations allow for entanglement purification \cite{gaussdistillability}.
However, it should be noted that the scheme presented here
endeavors to employ non-Gaussian operations in a minimal way,
namely only in the first step. Also, merely detectors are required
that distinguish the presence and absence of photons. 
The resulting procedure yields finally the desired output: the
states are approximately (i) pure, (ii) Gaussian, and (iii) highly
entangled. As such, this protocol could serve as a quantum privacy
amplification protocol in continuous-variable
\cite{PreskillCrypto,cryptography} entanglement-based quantum key
distribution. Needless to say, privacy amplification to pure
states is
only possible in case of an error-free implementation of the
required steps. In any practical realization, this is of course
not possible. One of the crucial numbers specifying to what extent
the procedure can achieve pure final states is the detection
efficiency of the used photon detectors. Yet, the purity
of the final states is no necessary requirement for 
quantum privacy amplification to function.

It should be noted that this procedure may well have significant
advantages compared to schemes in the finite-dimensional setting,
even when one is indifferent to whether the resulting states are
entangled in polarization or continuous degrees of freedom, and is
only interested in some form of highly entangled states of light.
The procedure is event-ready, in the sense that one has a
classical signal at hand indicating success, and in principle, one
can use the resulting states on the basis of this measurement. 
Photon counters, or set-ups where the photon number can be inferred
in retrospect from the outcomes of measurements on all modes including
the output modes, as well as controlled-not operations are not required.
Also, no post-processing is necessary dependent on the
measurement outcomes, which constitutes an advantage in iterative
protocols where the output of one step is the input of the next
step. But in turn, the final state that one obtains can obviously
not be the same maximally entangled state for all inputs, as for
finite-dimensional procedures this is often the case, one reason
being that the maximally entangled state does not even exist in
state space. There is no measure of the quality of the output
available as the fidelity with respect to a maximally entangled
state.

We have decided to split our analysis into several parts, in a
hierarchy of abstraction versus practicability: 
Following the introduction of the procedure in
Section \ref{procedure}, we will investigate the power of the
procedure in abstract terms. Section \ref{perfect} will be
dedicated to this analysis, where we will assume the applied
quantum operations to be error-free. The procedure can be
conceived as the continuous-variable analogue of the
finite-dimensional quantum privacy amplification procedure
\cite{Bennett}. We will present statements identifying the fixed
points of the iteration map: we will find that exactly the
Gaussians centred in phase space form the set of fixed points. We
will investigate convergence properties, and in particular present
necessary and sufficient criteria for convergence to a pure
entangled Gaussian state. The proofs -- some of them are 
technically involved -- will be presented in a
self-contained section in the Appendix. The degree of entanglement
and two-mode 
squeezing properties of the output will be investigated in
great detail, and several examples will be discussed. 

Section  \ref{PreparatoryStep} is concerned with the non-Gaussian
preparatory step in the absence and 
presence of decoherence. This step also relies only on 
photon detectors and passive linear optics.
In Section \ref{imperfect}
we will then discuss the situation when the performed 
operations are not
noiseless. The most relevant parameter here is the
detection efficiency of the photon detectors. We will present the
iteration map, and will investigate by numerical means what output
states should be expected asymptotically after many
iteration steps. 
In an experiment that realizes, say, a
single step of the distillation procedure, further sources of
imperfection play a role. This is the third part of the analysis.
These additional imperfections, including dark counts of the
detectors and mode matching issues will be sketched at the end of
Section \ref{imperfect} and will be discussed in great detail in a
forthcoming publication. Also, the quantum privacy amplification
capabilities in the presence of noisy apparata will be discussed
elsewhere. The present paper concentrates to a large extent to
flesh out what can be achieved in an asymptotic setting, or in a
protocol where many steps can be implemented: as such it shows
that continuous variable entanglement {\it can} be distilled, even
when resorting to the most feasible class of operations. 
For imperfect detectors, yet, it is shown that even after a small 
number of steps and for fairly low detection efficiencies,
the degree of entanglement can be significantly increased.
Finally,
in Section \ref{Summary} we summarise what we have achieved, and
sketch what further steps could be taken towards the
implementation of continuous-variable distillation schemes with
continuous variables.

\section{The procedure}\label{procedure}

The procedure makes use of only beam splitters and photon detectors
distinguishing the presence and absence of photons.
The iteration amounts to a very elementary operation, and
it will only turn out to be technical to characterize its properties
concerning convergence.
From a supply
of two-mode squeezed states, one prepares two-mode
 states $\rho$
in a first step.
The state $\rho$ is assumed to be non-Gaussian in the sense
that its associated characteristic function or Wigner function is not a
Gaussian in phase space.
This step can be conceived as the preparatory
part, this preparation
will be discussed in more detail in Section \ref{imperfect}.
Pairs $\rho\otimes \rho$
of such non-Gaussian states $\rho$
are then mixed at a $50:50$
beam splitter, yielding a state
\begin{eqnarray}
 (U\otimes U) (\rho\otimes \rho)(U\otimes U)^{\dagger},
\end{eqnarray}
where the beam splitters are reflected by unitaries \cite{vogelwelsch}
\begin{equation}
%
    U= T^{a_{1}^\dagger a_1} e^{-R^{\ast}a_{2}^{\dagger} a_{1} }
    e^{R a_{2} a_{1}^{\dagger}}
    T^{-a_{2}^\dagger a_2},
\end{equation}
where we set $T=R=1/\sqrt{2}$. Two of the output modes are then
directed into a photon detector, whose action is each
associated with Kraus operators
\begin{equation}
    E_1 = |0\rangle\langle 0|,\,\,  
    E_2 = {\mathbbm{1}}-|0\rangle\langle 0|,
\end{equation}
where $|0\rangle$ denotes the state vector
associated with the vacuum state.
Note that the first operator is a Gaussian projector and the second merely the difference of two Gaussian operators, which is a very simple non-Gaussian operation indeed.

\begin{figure}[th]
\centerline{
\includegraphics[width=8cm]{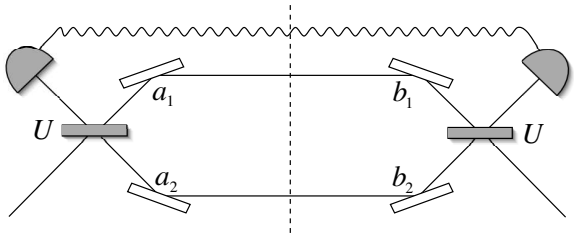}
}

\vspace{.2cm}
\caption{\label{fig:scheme1} A single step of the protocol: two non-Gaussian
states are mixed at a $50:50$ beam splitter. Two of the output
modes are fed into photon detectors, the remaining two modes are kept in further
steps, given a vacuum outcome in the detectors.}
\end{figure}

The state is kept in case of the vacuum
outcome of both local detectors. The procedure is
the Gaussification procedure of Ref.\ \cite{Gaussify}, but now the input states are not necessarily pure states.
The unnormalized final
state after one step is given by
\begin{eqnarray}
    \langle 0,0| (U\otimes U) (\rho\otimes \rho)(U\otimes U)^{\dagger}
    |0,0\rangle.
\end{eqnarray}
The resulting two-mode
states then form the basis of the next step. It is an iterative
protocol, and it is event-ready, in the sense that one has a
classical signal at hand which indicates
whether the procedure was successful or not. No further post-processing
has to be performed. The Gaussification
procedure is depicted in Fig.\ \ref{fig:scheme1}.

\section{Perfect devices}\label{perfect}

In this section we investigate the ideal situation where all
operations can be implemented to perfect accuracy. In particular,
perfect detection efficiencies are assumed, but later, this
assumption will be relaxed. We first discuss the recurrence
relation of the states, and then identify all fixed points of the
map. We then investigate convergence, and give a necessary and
sufficient criterion for convergence to a pure Gaussian two-mode
state. Note that the general situation that we encounter here is
different from the finite-dimensional setting. In the latter case,
the fidelity with respect to a maximally entangled state is a
meaningful measure of the quality of the output, for example the
fidelity with respect to the singlet state of two qubits. Here, no
such maximally entangled state exists, and consequently,
the degree of squeezing, of entanglement, and the von-Neumann entropy
characterizing purity are unbounded. Hence, a more complicated
interplay should be expected. We state the quality of the output
in terms of the degree of entanglement and squeezing of the
output.

\subsection{The iteration map}

The iteration map takes as input a four-mode state, and has
a two-mode state as output, which corresponds to the resulting state after the measurement,
given that both measurement outcomes were appropriate.
Let $\omega$ be a four-mode state; the iteration map ${\cal E}$
is then defined as
\begin{equation}
        {\cal E}(\omega)= \langle 0,0| (U\otimes U)  \omega
        (U\otimes U) ^\dagger |0,0\rangle.
\end{equation}
The output of one step is the input of the next step, which gives rise
to a sequence of states representing the output after a number of steps.
The map that links one state with the subsequent one -- given a positive outcome
of the measurements -- will be referred to as
recursion relation.

The recursion relation can be stated in a transparent way
as follows. The two-mode state $\rho$ can be represented
in the number
basis $\{|n\rangle:n\in{\mathbbm{N}}\}$ as
\begin{equation}\label{r1}
        \rho = \sum_{s,t,n,m=0}^\infty \rho_{s,t;n,m}
        |s,t\rangle\langle n,m|.
\end{equation}
The output $\rho'={\cal E}(\rho\otimes\rho)$ can then be written as
\begin{equation}\label{r2}
        \rho' = \sum_{s,t,n,m=0}^\infty \rho'_{s,t;n,m}
        |s,t\rangle\langle n,m|,
\end{equation}
where
\begin{eqnarray}
        \rho'_{a,b;c,d}  & = &
        \sum_{s=0}^a
        \sum_{t=0}^b
        \sum_{n=0}^c
        \sum_{m=0}^d
        M_{a,b;c,d}^{s,t;n,m} \nonumber\\
        &\times &
        \rho_{s,t;n,m} \rho_{a-s,b-t;c-n,d-m}.
\end{eqnarray}
The coefficients $M_{a,b;c,d}^{s,t;n,m}$ are given by
\begin{eqnarray}\label{nl1}
        M_{a,b;c,d}^{s,t;n,m} & =&
        2^{-(a+b +c+d)/2} (-1)^{(a+b+c+d)-(s+t+n+m)}\nonumber \\
        &\times &\left[
        \binom{a}{s}
        \binom{b}{t}
        \binom{c}{n}
        \binom{d}{m}
        \right]^{1/2}.
\end{eqnarray}
Note that each number $\rho'_{a,b;c,d}$ for some
$a,b,c,d\in{\mathbbm{N}}$ is a polynomial in $\rho_{s,t;n,m} $ for
$s\leq a$, $t\leq b$, $n\leq c$, and $m\leq d$ only. This is a
quite helpful property as it implies 
that changes of coefficients associated with certain Fock numbers
depend only on coefficients associated with smaller Fock 
numbers.
This property will prove to be crucial in the formal
proofs of convergence that are presented later on in this paper.

\subsection{Fixed points of the map}

The first task is to identify the fixed points of the iteration
map: these are the states that are left unchanged under the
operation. With fixed point we mean two-mode states $\rho$ for
which there exists a $c\in(0,1]$ such that
\begin{equation}
    \rho= c \; {\cal E}(\rho\otimes \rho).
\end{equation}
The number $c$ reflects the fact that we do not require the output
to be the appropriate one with unit probability. Some
fixed points can easily be found. These are all two-mode Gaussian
states with vanishing first moments.  That this set of
Gaussian states qualifies as being a set of fixed points becomes
obvious in terms of the covariance matrices.

This can be formulated as follows.
For a general $n$-mode system,
let $R=(X_{1},P_{1},\ldots,X_{n},P_{n})$ be
the vector consisting of the canonical coordinates of
an $n$-mode system (only two and four-mode systems will be
relevant). The canonical commutation relations (CCR)
$[R_{j},R_{k}] = i\Sigma_{j,k}$ for $j,k=1,\ldots,2n$
give rise to the symplectic matrix
\begin{equation}
    \Sigma = \bigoplus_{j=1}^{n}
    \left(
    \begin{array}{cc}
    0 & 1\\
    -1 & 0\\
    \end{array}
    \right).
\end{equation}
If the second moments exist, then the
covariance matrix $\gamma$ \cite{st}, a real symmetric
$2n\times 2n$-matrix, is defined as
\begin{eqnarray}
    \gamma_{j,k} =2 \text{Re}\left(\text{Tr}[
    (R_{j}-\text{Tr}[R_{j}\rho])
    (R_{k}-\text{Tr}[R_{k}\rho])
    \rho
    ]\right).
\end{eqnarray}
Every covariance matrix satisfies the Heisenberg uncertainty
principle $\gamma + i \Sigma \geq 0$.

Symplectic transformations \cite{st}
such as the one corresponding
to the application of the above beam splitters are reflected
by transformations of the form
\begin{equation}
    \gamma \longmapsto S\gamma S^{T}
\end{equation}
with $S\in Sp(2n,{\mathbbm{R}})$, i.e., $S\Sigma S^{T} =
\Sigma$ \cite{st}.
If the first moments vanish, the characteristic
function $\chi:{\mathbbm{R}}^{2n}\longrightarrow {\mathbbm{C}}$
defined as $\chi(\xi)=\text{tr}[W_\xi \rho]$
of a Gaussian state $\rho$
can be written as
\begin{equation}
    \chi(\xi)= 
e^{-(\xi^{T} \Sigma^{T}\gamma \Sigma \xi)/4},
\end{equation}
where the Weyl displacement operator is given by
\begin{equation}
    W_{\xi}= \exp( i \xi^{T} \Sigma R).
\end{equation}
$\gamma$ is just the above covariance matrix.
With this notation, the fact that Gaussians in
the centre of phase space form fixed points becomes
straightforward.

\begin{proposition}[Gaussian states are fixed points]{Let
$\rho$ be a Gaussian two-mode state with vanishing first moments.
Then ${\cal E}(\rho\otimes \rho)=\rho$.\label{Lem}
}
\end{proposition}

In fact, this set of fixed points is already exhaustive: there are
no other fixed points that do not have the property of Proposition
\ref{Lem}. This is quite a surprise, but a pleasant one, as it
makes the characterisation of the fixed points very easy: the
fixed points are exactly the Gaussian states with vanishing first
moments, and there are no other states that are left unaffected by
the iteration. The proof of this assertion can be found in Appendix \ref{appa}.

\begin{proposition}[All fixed points are Gaussian]{\label{allgauss}
Let $\rho$ be a two-mode state satisfying $c\; {\cal
E}(\rho\otimes \rho)=\rho$ for some $c>0$. Then $\rho$ is a
Gaussian state with vanishing first moments.}
\end{proposition}

The proof of this statement -- while being of central interest for
the purposes of this paper -- is quite technical and will for the
sake of readability of the main text be presented in Appendix \ref{appb}.
The property stated in Proposition 2 is an interesting feature of
the iteration map: it is a Gaussian map and hence maps by
definition all Gaussian states onto Gaussian states. But even when
applied to general two-mode quantum states, the only fixed points
of the map happen to be Gaussian states as well. No other state is
left unchanged under the mixing at the beam splitters followed by
the measurement.

\subsection{Convergence}

Once the fixed points have been specified, the next step is to see
whether the procedure in fact converges to such a Gaussian state
with vanishing first moments. So the question is under what
constraints the procedure converges to a state at all -- which is
then one of the fixed points. We will also
characterize fully those initial states, pure or mixed, for which
convergence can be shown to an exactly pure Gaussian state.

The above protocol is probabilistic, and in the
following convergence proofs it is assumed that it is
successful in each step. This is to demonstrate that as a matter
of principle, this procedure has the power of distilling entangled
Gaussian states from a supply of non-Gaussian initial states.
The closer the state approaches the fixed point, the
smaller the improvement in quality in a step of our procedure. As
every step of the procedure loses at least half of the partially
entangled pairs that enter the iteration step one has to make a
meaningful choice concerning the accuracy to which the envisioned
entangled state can be achieved on the one hand and the rate with
which this is possible on average on the other hand. 

For a $\rho\in{\cal S}$, where ${\cal S}$ denotes the state space
on ${\cal H}\otimes {\cal H}$, let ${\cal E}^{(1)}(\rho)$ be
defined as
\begin{equation}
    {\cal E}^{(1)}(\rho) = {\cal E}(\rho\otimes \rho),
\end{equation}
and
\begin{eqnarray}
    {\cal E}^{(i+1)}(\rho) = {\cal E}({\cal E}^{(i)}(\rho)\otimes {\cal E}^{(i)}(\rho))
\end{eqnarray}
for $i\in{\mathbbm{N}}$. So ${\cal E}^{(i+1)}(\rho)$ is nothing
than the unnormalized outcome of the procedure after $i+1$ steps
of the iteration. The actual states are then
\begin{equation}
    \rho^{(i)}= \frac{{\cal E}^{(i)}}{\text{tr}[ {\cal E}^{(i)}]}.
\end{equation}
The next result states when convergence to one of the
fixed points can be expected.

To express the result in a concise manner, introduce a useful map,  
$B:{\cal S}\longrightarrow
M_{4\times 4}$, which for any state  $\rho$ allows one to 
construct a real symmetric $4\times4$ matrix $B(\rho)$, with which we will 
characterize the convergence properties of the protocol.
The map $B$ can be defined in terms of the matrix 
elements of $B(\rho)$ as follows:
\begin{eqnarray}
B(\rho)_{1,1}&=&\frac{1}{2}\left(
-(\sigma_{1,0,1,0}-1)+\sqrt{2}\real(\sigma_{2,0,0,0})\right)\nonumber\\
B(\rho)_{2,2}&=&\frac{1}{2}\left(
-(\sigma_{1,0,1,0}-1)-\sqrt{2}\real(\sigma_{2,0,0,0})\right)\nonumber\\
B(\rho)_{3,3}&=&\frac{1}{2}\left(
-(\sigma_{0,1,0,1}-1)+\sqrt{2}\real(\sigma_{0,2,0,0})\right)\nonumber\\
B(\rho)_{4,4}&=&\frac{1}{2}\left(
-(\sigma_{0,1,0,1}-1)-\sqrt{2}\real(\sigma_{0,2,0,0})\right)\nonumber\\
B(\rho)_{1,2}&=&\frac{1}{\sqrt{2}}\imag(\sigma_{2,0,0,0})\nonumber\\
B(\rho)_{3,4}&=&\frac{1}{\sqrt{2}}\imag(\sigma_{0,2,0,0})\nonumber\\
B(\rho)_{1,3}&=&\frac{1}{2}\left(
-\real(\sigma_{1,0,0,1})+\real(\sigma_{1,1,0,0})\right)\nonumber\\
B(\rho)_{1,4}&=&\frac{1}{2}\left(
\imag(\sigma_{1,0,0,1})+\imag(\sigma_{1,1,0,0})\right)\nonumber\\
B(\rho)_{2,3}&=&\frac{1}{2}\left(
-\imag(\sigma_{1,0,0,1})+\imag(\sigma_{1,1,0,0})\right)\nonumber\\
B(\rho)_{2,4}&=&\frac{1}{2}\left(
-\real(\sigma_{1,0,0,1})-\real(\sigma_{1,1,0,0})\right)\label{superlong}
\end{eqnarray}
where we define $\sigma=
\rho/\rho_{0,0,0,0}$, which is the state $\rho$ normalized such that element $\sigma_{0,0,0,0}=1$
for  states
$\rho\in {\cal S}$ with $\rho_{0,0,0,0}=\langle 0,0| \rho
|0,0\rangle>0$.
 The map is not defined for states with $\rho_{0,0,0,0}=0$, but that has no 
consequences for this discussion.

 For a Gaussian two-mode state $\rho$, the $4\times
4$-matrix $B(\rho)$ is always invertible, and its covariance matrix is simply
\begin{equation}
    \gamma= \Sigma^{T}B(\rho)^{-1}\Sigma+ 
	{\mathbbm{1}}_{4}.
\end{equation}

\begin{proposition}[Convergence to Gaussian states]{\label{p2}
Let $\rho\in{\cal S}$. Then there exists a Gaussian two-mode state
with vanishing first moments $\omega$ satisfying
\begin{eqnarray}
\lim_{i\rightarrow \infty}
\langle a,b| \left(\rho^{(i)} -
\omega \right) |c,d\rangle=0
\end{eqnarray}
for all $a,b,c,d\in {\mathbbm{N}}_{0}$ if and only if
both (i) $\langle 0,0
    |
    {\cal E}(\rho\otimes \rho)
    |0,0\rangle >0$,
and (ii) $B(\rho^{(1)})$ is invertible.
}

\end{proposition}

The matrix 
\begin{equation}
\Sigma^T B(\rho^{(1)})^{-1}\Sigma 
+ {\mathbbm{1}}_{4}
\end{equation}
 with
$\rho^{(1)}= {\cal E}(\rho\otimes \rho)/\text{tr} [{\cal
E}(\rho\otimes \rho)]$ is then the covariance matrix of the
resulting Gaussian state to which the procedure converges (weakly).
In other words, the procedure always converges to Gaussian two-mode
states, whenever there is a non-vanishing success probability.
The proof of this statement will be presented in 
Appendix \ref{appc}. The most interesting case is the situation where one can
prove convergence to an exactly pure Gaussian state. This is the
case when $\Sigma^T B(\rho^{(1)})^{-1}\Sigma
+ {\mathbbm{1}}_{4}$ is the
covariance
matrix of a pure Gaussian state. In this case, however, the
necessary and sufficient criteria can be formulated in a more
transparent manner. For imperfect detectors one will not achieve
this goal exactly. From the perspective of the basic understanding
of the procedure, however, necessary and sufficient conditions for
convergence are instructive.

\begin{proposition} {\bf (Necessary and sufficient conditions
    for convergence to pure Gaussians)}\label{PureProp}
Let $\rho\in {\cal S}$.
Then there exists a pure
two-mode Gaussian state $|\psi\rangle\langle\psi|$
centred in phase space satisfying
\begin{eqnarray}
\lim_{i\rightarrow \infty}
\langle a,b| \left(\rho^{(i)} -
|\psi\rangle\langle\psi|
\right) |c,d\rangle=0
\end{eqnarray}
for all $a,b,c,d\in {\mathbbm{N}}_{0}$
if and only if (i) $\rho_{0,0,0,0}>0$, (ii)
\begin{eqnarray}\label{condi}
        \rho_{1,0,1,0}&=& |\rho_{1,0,0,0}|^2/\rho_{0,0,0,0},\\
        \rho_{0,1,0,1}&=& |\rho_{0,1,0,0}|^2/\rho_{0,0,0,0},\\
        \rho_{1,0,0,1}&=& |\rho_{1,0,0,0}|^2/\rho_{0,0,0,0},
\end{eqnarray}
and (iii)
\begin{eqnarray}
   \left\|
   \begin{array}{cc}
       \sqrt{2} \frac{\rho_{2,0,0,0}}{\rho_{0,0,0,0}}
       - \frac{\rho_{1,0,0,0}^{2}}{\rho_{0,0,0,0}^{2}} &
       \frac{\rho_{1,1,0,0}}{\rho_{0,0,0,0}}
       -\frac{\rho_{1,0,0,0}\rho_{0,1,0,0}}{\rho_{0,0,0,0}}\\
       \frac{\rho_{1,1,0,0}}{\rho_{0,0,0,0}}
       -\frac{\rho_{1,0,0,0}\rho_{0,1,0,0}}{\rho_{0,0,0,0}}
       &
       \sqrt{2} \frac{\rho_{0,2,0,0}}{\rho_{0,0,0,0}} -
       \frac{\rho_{1,0,0,0}^{2}}{\rho_{0,0,0,0}^{2}}
   \end{array}
   \right\|_{\infty}<1,\nonumber\\
\end{eqnarray}
where $\|.\|_{\infty}$ denotes the spectral norm.
\end{proposition}
The proof of Proposition 4 can be found in Appendix~\ref{appd}.
In this sense
the protocol can also work as a purification
protocol. Not only the degree of entanglement and
squeezing are increased, but also the von-Neumann
entropy decreased. Note, however, that the conditions
to convergence to exactly pure Gaussian states are
fairly restrictive, and one will often encounter the situation
that the states are almost pure. It can also happen that
the degree of entanglement is significantly
increased in the course of the procedure,
but the von-Neumann entropy also increases.

Such a procedure would -- if implemented with perfect physical
devices -- clearly lead to a perfect distillation procedure, and
moreover, to the possibility of
quantum privacy amplification for unconditionally
secure quantum key distribution:
if the state is a pure state, it cannot be entangled to any system of
a third party, and the resulting states could be used to simply
transmit a quantum state encoding the key. Needless to say, one would
still have to apply classical methods to extract a secure key from
the measurement outcomes, as the states exhibit only a finite
squeezing, and hence, measurements of canonical coordinates by means
of homodyne detection do not yield a deterministic outcome.
This is, however,
then entirely due to the Gaussian character of the quantum
state and not due to an eavesdropper.

Note that the second moments of the states do not remain constant
during this procedure, as they are modified in the steps involving
measurements. For example, the only states with the second moments
of a Gaussian pure state is a Gaussian state itself.
Hence, the `Gaussification' of the states is not
merely a mixing to a Gaussian state, by virtue of the law of large
numbers.

\subsection{Examples}

We will now turn to discussing a number of examples to visualize
the procedure.  For a number of initial states we
will discuss the degree of entanglement after a single
step and
compare it with the degree of entanglement of
the Gaussian state the sequence of states converges to.
We start by giving an example for pure states, and
then turn to mixed initial states.\\

\begin{figure}[th]
\centerline{
\includegraphics[width=7.5cm]{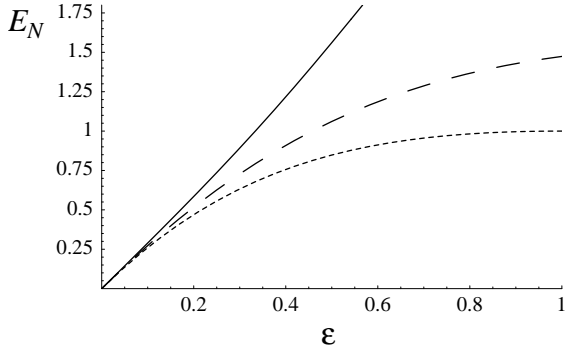}
}

\vspace{.2cm}
\caption{\label{fig:pure1} The logarithmic negativity of the resulting states
for an initial state $\rho$ as specified in Eq.\ (\ref{spec}). Depicted is
the logarithmic negativity of the Gaussian state the sequence converges to (straight
line), i.e., the degree of entanglement after 'infinitely many steps', (ii)
of the output state after a single iteration of the protocol (dashed line), and (iii)
of the initial state $\rho$ itself (dotted line). Note that the increase of entanglement
is already significant in a single step of the procedure. For infinitely many iterations and
$\varepsilon\rightarrow 1$ the degree of entanglement diverges.}
\end{figure}

{\it Example 1:} Let $\rho\in {\cal S}$ be the (pure) state with
\begin{eqnarray}\label{spec}
    \rho_{0,0,0,0}&=&1/(1+\varepsilon^{2}),\\
    \rho_{1,1,0,0}& =&\rho_{0,0,1,1}=\varepsilon/(1+\varepsilon^{2}),\nonumber \\
    \rho_{1,1,1,1}&=& \varepsilon^{2}/(1+\varepsilon^{2}),\nonumber
\end{eqnarray}
with $\varepsilon\in [0,1)$, 
and $\rho_{a,b,c,d}=0$ with $a,b,c,d\in
{\mathbbm{N}}_0$ otherwise. The sequence of outputs converges to a
Gaussian pure state, the covariance matrix of which can be
evaluated according to Proposition \ref{p2}. Depending on the
value of $\varepsilon$, the degree of entanglement that can be
achieved will be different. This degree of entanglement will now
and in the following be quantified in terms of the logarithmic
negativity \cite{logneg}, which is for a state $\rho$ of a
bi-partite system defined as
\begin{equation}
    E_{N}(\rho) = \log_{2}\| \rho^{\Gamma}\|_{1},
\end{equation}
where $\|.\|_{1}$ denotes the trace-norm, and
$\rho^{\Gamma}$ is the partial transpose of $\rho$ \cite{QuantNote}.
The negativity has been shown to be monotone under
local operations with classical communication \cite{lognegmon,lognegmon2}.
The logarithmic negativity is an upper bound for the
distillable entanglement, grasping the resource character
of entanglement. For Gaussian states, as we encounter here,
the logarithmic
negativity has moreover a clearcut physical interpretation
in terms of an asymptotic cost function \cite{PPT}.
For the specific initial state  as in Eq.\ (\ref{spec}),
Fig.\ \ref{fig:pure1} displays the behavior the logarithmic negativity
of the initial state $\rho$, and of the Gaussian state the
sequence of states converges to
as a function of $\varepsilon \in[0,1)$.
Note that the corresponding expressions
can be evaluated analytically, by virtue of Proposition \ref{p2}.

For $\varepsilon\rightarrow 1$ the logarithmic negativity of the resulting
Gaussian state diverges, which reflects the fact that with unit
detection efficiencies, one could in principle prepare arbitrarily
entangled and squeezed states in this manner. This holds, as
should be noted, also in the single-mode case, and one may prepare
states with an arbitrary degree of single-mode squeezing.
Technical limitations, however, will limit the degree of squeezing
that is actually attainable, as will be discussed later.

\begin{figure}[th]
\centerline{
\includegraphics[width=7.5cm]{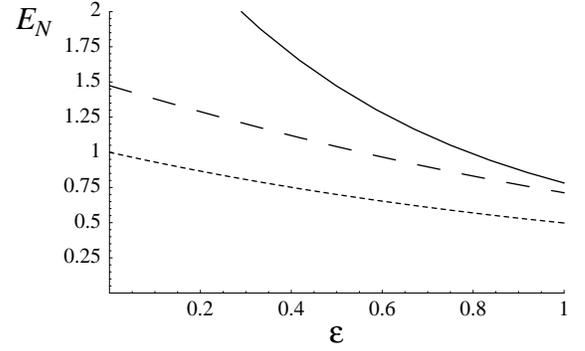}
}

\vspace{.2cm}
\caption{\label{fig:pure2} The logarithmic negativity of the resulting states
for initial states $\rho$ as in Eq.\ (\ref{spec2}). Again, the straight line
shows the negativity of the Gaussian state the sequence converges to,
with diverging logarithmic negativity as $\varepsilon$ approaches zero. The dashed
line represents the logarithmic negativity after a single step, whereas the dotted line
is the logarithmic negativity of the original state. }
\end{figure}

{\it Example 2:} As another example, let us consider the state
$\rho\in {\cal S}$  with
\begin{eqnarray}\label{spec2}
    \rho_{0,0,0,0}&=&\rho_{1,1,0,0} =\rho_{0,0,1,1} = \rho_{1,1,1,1}\nonumber\\
    &=&
    1/(2+\varepsilon),\\
    \rho_{0,1,0,1}&=& \varepsilon/(2+\epsilon)
    \nonumber
\end{eqnarray}
with $\varepsilon\in [0,\infty)$,
and $\rho_{a,b,c,d}=0$ otherwise. This state is mixed
for all values $\varepsilon\in (0,\infty)$. Fig.\ \ref{fig:pure2}
depicts again the value of the logarithmic negativity of the initial state,
after a single iteration step and after 'infinitely many' iterations.

\begin{figure}[th]
\centerline{
\includegraphics[width=7.5cm]{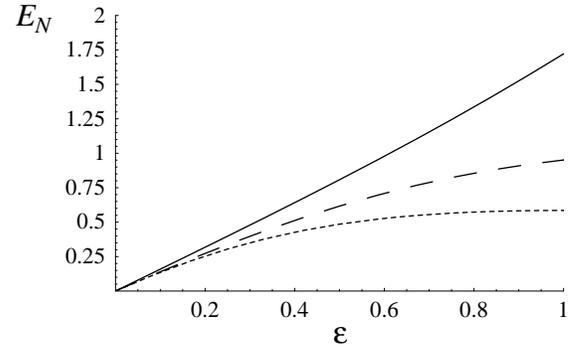}
}

\vspace{.2cm}
\caption{\label{fig:pure3} As Fig.\ \ref{fig:pure2}
for the initial state as in Eq.\ (\ref{spec3}).}
\end{figure}

\begin{figure}[th]
\centerline{
\includegraphics[width=7.5cm]{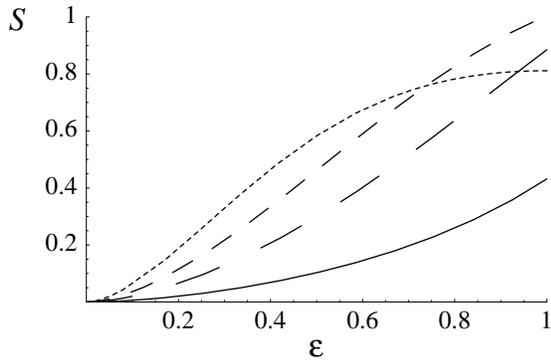}
}

\vspace{.2cm}
\caption{\label{fig:pure4} This figure depicts
the von-Neumann
entropy of the output as a function of
$\varepsilon\in [0,1)$, for
the initial state as in Eq.\ (\ref{spec3}).
The dotted line corresponds to the initial state itself,
and the dashed lines show the von-Neumann entropy
after one (dashed), two (dashed with longer
dashing), and four steps (solid line).
The limiting
von-Neumann entropy vanishes exactly,
as the conditions for
Proposition \ref{PureProp} are met. Note that for
very mixed initial states the von-Neumann entropy
can first increase, to then decrease again
in further steps.}
\end{figure}

{\it Example 3:} The third example is
$\rho\in {\cal S}$  with
\begin{eqnarray}
    \rho_{0,0,0,0}&=&1/(1+\varepsilon^2),
    \,\,
    \rho_{1,1,0,0} =\rho_{0,0,1,1} = \varepsilon/(2+2\varepsilon^2)\nonumber\\
     \rho_{1,1,1,1}&=&\varepsilon^2/(1+\varepsilon^2)\label{spec3}
\end{eqnarray}
with $\varepsilon\in [0,1)$, and $\rho_{a,b,c,d}=0$ otherwise.
Again, we depict the degree of entanglement in Fig.\
\ref{fig:pure3}. This is a set of initial states for which exact
convergence to a pure Gaussian state is guaranteed by Proposition
\ref{PureProp} for all values of $\varepsilon$. In Fig.\
\ref{fig:pure4}, the von-Neumann entropy $S(\omega)=-
\text{tr}[\omega\log_2\omega]$ for states $\omega$ is displayed,
again as a function of $\varepsilon$, characterizing the purity of
the resulting states. Note that the state does not only become
more entangled, but also less mixed in the course of the protocol.

\begin{figure}[th]
\centerline{
\includegraphics[width=6.5cm]{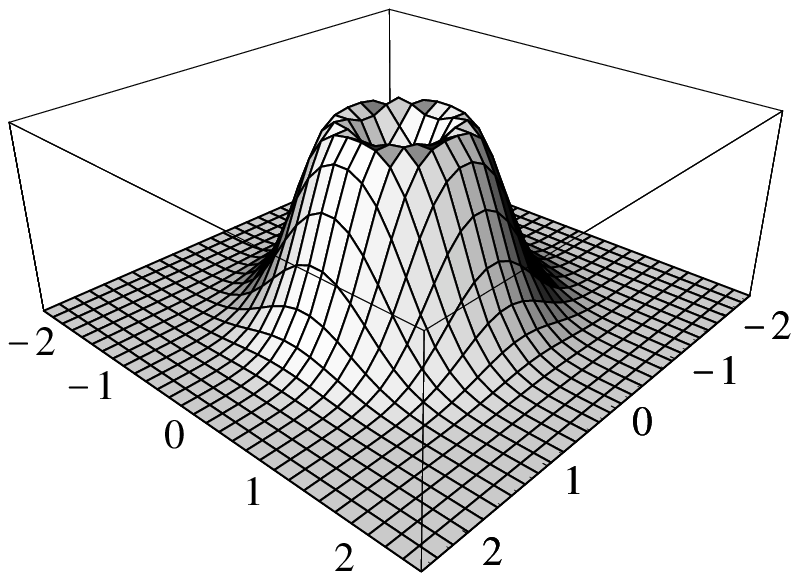}
}
\centerline{
\includegraphics[width=6.5cm]{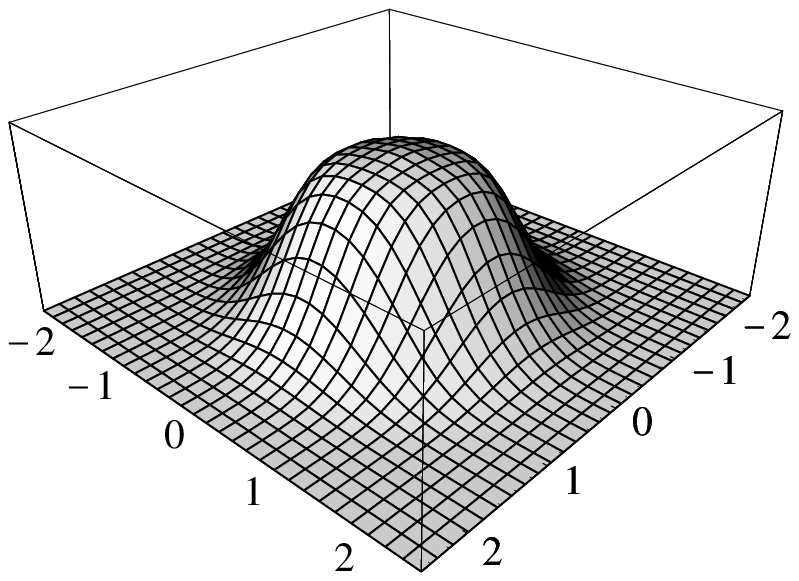}
}
\centerline{
\includegraphics[width=6.5cm]{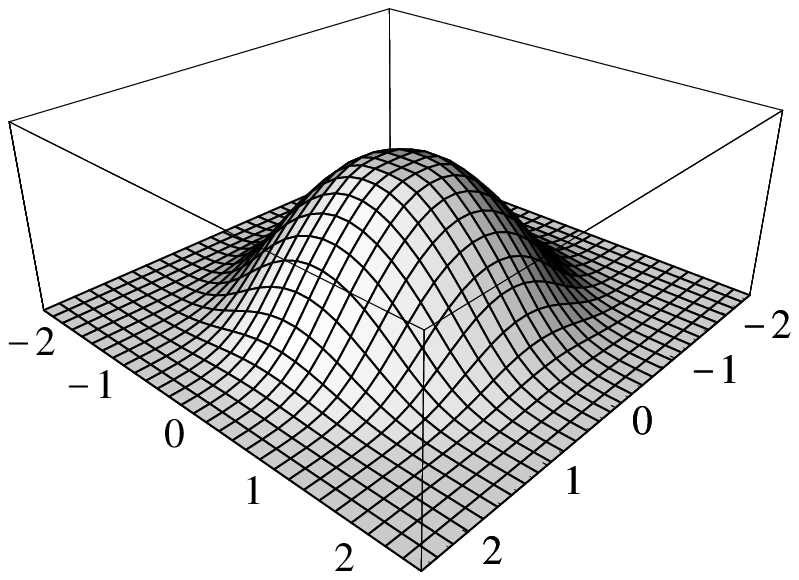}
}

\vspace{.2cm}
\caption{\label{fig:Wigner} The Wigner function of the reduced
state of  one system initially and after one and two steps for the
initial state as in Eq.\ (\ref{spec}).}
\end{figure}

{\it Example 4:} As a mere graphical
illustration of the procedure
we represent the states that are encountered in the procedure
in phase space. More specifically, for the initial state $\rho$
as in Example 1 with $\varepsilon=0.6$ we investigate the
Wigner function of the single-mode states of the reduced
state of one mode alone for the initial state, and the state
after one and two steps.
The Wigner function $W:{\mathbbm{R}}^{2}\longrightarrow
{\mathbbm{R}}$ is the Fourier
transform of the characteristic function,
\begin{equation}
        W(\xi) = \frac{1}{(2\pi)^2} \int e^{i \xi^T \Sigma \eta} \chi(\eta) d^2 \eta.
\end{equation}
Fig.\ \ref{fig:Wigner} shows the Wigner function of
$\text{tr}_B[ \rho]$, $\text{tr}_B[ \rho^{(1)}]$, and
$\text{tr}_B[ \rho^{(2)}]$. While initially, the Wigner function is
far from being a Gaussian in phase space, its non-Gaussian features
disappear quickly.

\subsection{Achievable squeezing}

Another relevant
figure is the degree of two-mode squeezing that one
can achieve. For a Gaussian state $\rho$ with covariance matrix $\gamma$,
one of the standard definitions for the degree of squeezing translates into
the language of covariance matrices to
\begin{equation}
        E_S(\rho) = \max
	\left\{- \log
\lambda_{\text{min}}(\gamma),0\right\}
\end{equation}
where $\lambda_{\text{min}}$ denotes the smallest eigenvalue.
Roughly speaking,
one compares the variance with respect to
some canonical coordinate of this state
with the one of the vacuum.
For example, for the two-mode squeezed state $\rho$
with a covariance matrix
of the form (\ref{cvf}) with
\begin{equation}\label{squeezing}
        \xi= \cosh(2r),\,\,
        \zeta = \sinh(2r),
\end{equation}
$r\in[0,\infty)$ being the ordinary two-mode squeezing parameter,
we arrive at $E_S(\rho)=r$. In terms of  dB,
the degree of squeezing is related to $E_S(\rho)$
according to
$
2 E_S(\rho)/\log(10)$.
For a two-mode state, this gives the
total degree of squeezing. To distinguish genuine two-mode squeezing
from locally available single-mode squeezing, one may consider
for a two-mode state with $4\times 4$ covariance matrix of
the block
form
\begin{equation}
        \gamma=\left(
        \begin{array}{cc}
        \gamma_A & \gamma_C\\
        \gamma_C^T & \gamma_B
        \end{array}
        \right)
\end{equation}
the quantity
\begin{eqnarray}
        E_{TS}(\rho) = \max\left\{- \log \lambda_{\text{min}}
	((S_A\oplus S_B) \gamma(S_A\oplus S_B)^T),0\right\},
        \nonumber \\
\end{eqnarray}
where $S_A,S_B\in Sp(2,{\mathbbm{R}})$ are chosen in such a way such that
\begin{equation}
        S_A \gamma_A S_A^T=\alpha {\mathbbm{1}}_2,\,\,\,
        S_B \gamma_B S_B^T=\beta {\mathbbm{1}}_2
\end{equation}
with $\alpha,\beta\geq1$, and $S_A \gamma_C S_B^T  $ is diagonal. This is the degree
of squeezing after the state has been locally brought to a 
form corresponding to a Gibbs state with
appropriate local symplectic transformations 
(that can in particular include also single-mode squeezing).

\section{Preparatory step}\label{PreparatoryStep}

The above procedure takes non-Gaussian states as input. If one starts
off with a Gaussian state, 
a
preparatory step is clearly necessary. Such a step has been described in
Ref.\ \cite{Gaussify}. In principle, any other protocol preparing the
states that are suitable in the above procedure would be appropriate as well.
The one presented here has the advantage that it does not leave the scope
of the paper, in that only the possibility of preparing
Gaussian states is assumed, and the use of dichotomic photon detectors
with high efficiency, which do not resolve the photon number.

\subsection{The preparation procedure}

\begin{figure}[]
\centerline{
\includegraphics[width=9cm]{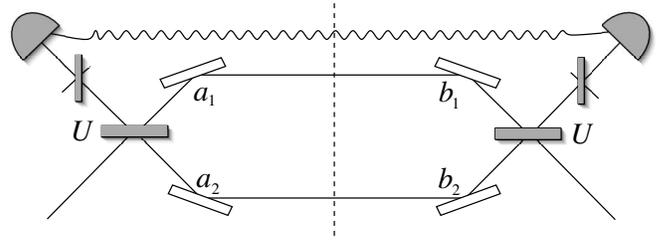}
       }

\vspace{.2cm} \caption{\label{fig:scheme2} A single step of the
protocol for detection efficiencies smaller than one. The
detection efficiency less than unity is incorporated by placing
perfect detectors behind a beam splitter.}
\end{figure}

Let us assume that one has two-mode squeezed states at hand, which are
Gaussian states the covariance matrix of which is of the form
\begin{equation}\label{cvf}
        \gamma=\left(
        \begin{array}{cccc}
        \xi & 0 & \zeta & 0\\
        0 & \xi & 0 & -\zeta\\
          \zeta & 0 &  \xi & 0\\
          0 & -\zeta & 0 & \xi
        \end{array}
        \right),
\end{equation}
where $\zeta^2= \xi^2-1$, $\xi \in (0,\infty]$. The numbers
$\xi,\zeta$ characterize the degree of
two-mode squeezing, as in Eq.\ (\ref{squeezing}).
It has been shown in Ref.\ \cite{Gaussify} that with
the procedure as in Fig.\ \ref{fig:scheme2},
with the appropriate choice for $\xi$ and for the transmittivity $T$ and reflectivity  $R$
of the beam splitter $V$,
the resulting mixed two-mode state can be made arbitrarily close
in trace-norm to a maximally entangled state
in the subspace isomorphic to ${\mathbbm{C}}^2\otimes {\mathbbm{C}}^2$
 with state vector $|0,0\rangle + \lambda |1,1\rangle/(1+\lambda^2)^{1/2}$ for
 any value of $\lambda\in [0,1)$.
This state can be used as input of the procedure. Taking this as input of the procedure
of Section \ref{procedure}, the sequence of states will converge to a pure state
with an arbitrarily large
degree of entanglement, as well as an arbitrary degree of two-mode squeezing.

\begin{figure}[]
\centerline{
\includegraphics[width=8cm]{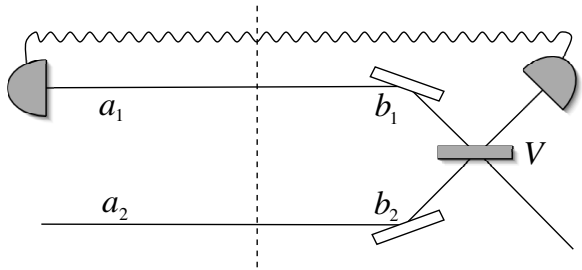}
       }

\vspace{.2cm} \caption{\label{fig:scheme3} The preparatory step
which resembles very much the step of an iteration in the
Gaussification scheme. Given two specimens of two-mode squeezed
states, measurements are performed on two of the four modes.
However, now the outcome is accepted in case that both detectors
click, corresponding to the Kraus operators $E_2
={\mathbbm{1}}-|0\rangle\langle0|$. The beam splitter $V$ is not a
$50:50$ beam splitter, but one that has an appropriately tuned
reflectivity and transmittivity.}
\end{figure}

\subsection{The full protocol under decoherence}

Yet,
such states are not available in the presence of a decoherence process. Once locally prepared, two-mode squeezed
states or other entangled Gaussian states will
deteriorate into weakly entangled mixed states.
The decoherence dynamics we are interested in, i.e., 
transmission through
an absorbing fiber, does not alter the Gaussian character 
of the state \cite{scheelwelsch}, and it acts in effect
as a Gaussian channel \cite{Channel}.
We assume that both modes are affected in the same
manner.

\begin{figure}[]
\centerline{
\includegraphics[width=8cm]{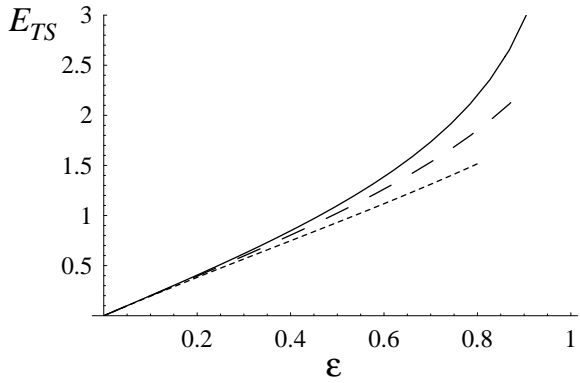}
       }

\vspace{.2cm}
\caption{\label{fig:squeezing} The degree of two-mode squeezing
for the Gaussian to which convergence is encountered, as a function of
$\lambda\in[0,1)$, for $\theta=1$ (solid line), $\theta=0.9$ (dashed line),
and $\theta=0.8$ (dotted line).}
\end{figure}

\begin{figure}[]
\centerline{
\includegraphics[width=8cm]{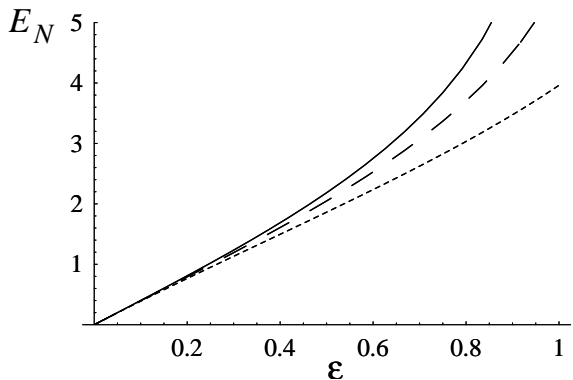}
       }

\vspace{.2cm}
\caption{\label{fig:entanglement} As Fig.\ \ref{fig:squeezing}, but
for the degree of entanglement in terms of 
the achievable logarithmic negativity.}
\end{figure}

Starting from the two-mode state with covariance matrix $\gamma$ 
as in Eq.\ (\ref{cvf}),
the state after the decoherence process is a Gaussian state with a covariance matrix of
the same simple form, but with the roles of $\xi$ and $\zeta$ being replaced by
\begin{equation}
        \xi\longmapsto 1+\theta^2 (\xi-1),\,\,\,\,
        \zeta\longmapsto \theta^2 \zeta,
\end{equation}
$\theta\in [0,1]$ specifying the strength of the decoherence
process. The value $\theta=1$ corresponds to no losses at all. It
is a tedious but 
straightforward calculation to show (compare, e.g., Ref.\
\cite{Nonlin}) that for each value of $\theta$, one can find a
choice for $\xi$ and the transmittivity $T$ and reflectivity $R$
such that the resulting state becomes arbitrarily close in
trace-norm to $\rho\in {\cal S}$ with
 \begin{eqnarray}
 \rho_{0,0,0,0}&=& 1/(1+ \lambda^2\varepsilon + \lambda^2 ),\\
 \rho_{1,1,0,0} &=&\rho_{0,0,1,1} = \lambda \rho_{0,0,0,0},\\
 \rho_{1,1,1,1} &=  &  \lambda^2 \rho_{0,0,0,0},\\
\rho_{0,1,0,1} &= &\varepsilon\lambda^2 \rho_{0,0,0,0},
\end{eqnarray}
with
\begin{equation}
        \varepsilon= (1-\theta^2)/\theta^2,
\end{equation}
and $\rho_{a,b,c,d}=0$ otherwise. Any value for the number $\lambda\in[0,1]$
can be achieved. For $\lambda=1$
this is nothing but the
state of Example 2 given in Eq.\ (\ref{spec2}).
The negativity after a number of steps of
the Gaussification protocol for this initial state
has already been depicted in Fig.\ \ref{fig:pure2}.
If one implements the full procedure, including the
preparatory step and one or
several instances of the whole procedure, it is
certainly the relevant figure to
compare the resulting degree
of entanglement with the one
before even the preparatory step has been
taken. However, the
initial two-mode squeezed states that have to be transmitted
before the preparatory step has a negligible degree of
entanglement: in order to better and better
approximate the state of the form as in Eq. (\ref{spec2}), less
and less entanglement is needed, at the expense that the achievable
rate decreases. Therefore, in principle, with perfect detection efficiencies
as considered in this section, one can encounter an arbitrarily large
entanglement gain. Also, note that the largest increase in entanglement
per step
occurs already in the preparatory step -- this is specific for this case of
perfect detectors.
Figs.\ \ref{fig:squeezing} and \ref{fig:entanglement} 
show the degree
of squeezing and entanglement that can be asymptotically achieved
as a function of $\lambda$ for several values of $\theta\in [0,1]$.

\section{Imperfect devices}\label{imperfect}

Until now, it was assumed that the devices that are being used are
noiseless and error free. We have seen that with such devices,
distillation to highly entangled Gaussian states is indeed
possible. This is the continuous-variable analogue to quantum
privacy amplification in the finite-dimensional case. In any
practical realization of such a scheme, however --  just as in the
case of finite systems such as qubits -- the performance of the
scheme depends very much on to what accuracy the operations can be
performed. The most important number here is the detector
efficiency of the photon detectors.

Needless to say, there are other potential sources of error: there
is no intrinsic 
need for storing light in fiber loops, but nevertheless,
losses even in short fibers cannot be avoided. If the scheme is
not implemented entirely within fibers, losses are due to the
coupling into the fiber. Then, in the entire paper we have
discussed single modes only, and not broadband squeezed states
\cite{TMSSTheory}. In a practical realization, mode-matching
problems will occur. Finally, dark counts of the photon detectors
must be included in the treatment: here, however, it turns out
that dark counts are much less problematic as one is tempted to
think on intuitive grounds. Firstly, in the Gaussification step
the only effect of dark counts is that one accidently does not
accept the resulting state. This merely leads to a reduced rate of
production of entangled states, but does not result in an increase
of noise in the outputs. As the whole protocol is assumed to be
implemented with a high repetition rate, this reduction of rate is
acceptable. In the preparatory step dark counts do play a role.
But even here, as the preparation requires a coincidence of
several or many detectors in a very short time window (four in the
simplest case), the dark count rate can in principle be made
arbitrarily small. These further sources of errors will be
discussed in a forthcoming paper. As has been pointed out in the
introduction, in this paper, we concentrate on the most crucial
source of error, the non-unit detection efficiency.

\subsection{The iteration map}

The case of imperfect detectors is modeled as follows, the 
usual projection operator  onto the 
number 
state $|n\rangle\langle n|$ is replaced by 
\begin{equation}
|n\rangle\langle n|\longmapsto 
\sum_{k=n}^\infty
{k \choose n}\eta^n
(1-\eta)^{k-n}|k\rangle\langle k|,
\end{equation}
with $\eta\in[0,1]$. This formula
has a representation in terms of a beam-splitter of 
transmittivity $T=\sqrt{\eta}$ placed in front of a detector 
with unit efficiency (see, e.g.,
Ref.\ \cite{K}).
As before, let $\rho$ be some trace-class
two-mode positive operator, i.e., a not necessarily
normalized two-mode state.
In the Fock basis, it can be represented
as in Eq.\ (\ref{r1}).
The resulting $\rho'$ after one step of the
procedure,
\begin{equation}
\rho'={\cal E}_T (\rho\otimes\rho),
\end{equation}
where ${\cal E}$ is replaced by the map  ${\cal E}_{T}$
reflecting non-unit efficiency,
can then be written as
\begin{equation}\label{rr2}
        \rho' = \sum_{s,t,n,m=0}^\infty \rho'_{s,t;n,m}
        |s,t\rangle\langle n,m|.
\end{equation}
The new coefficients can be evaluated to be given by
\begin{eqnarray}
        &&\rho'_{A,B;C,D}   =
        \sum_{k,l=0}^\infty
        \sum_{a=0}^{A+k}
        \sum_{b=0}^{B+l}
        \sum_{c=0}^{C+k}
        \sum_{d=0}^{D+l}\\
&\times &
        \sum_{s=0}^{\min\{a,k\}}
        \sum_{u=0}^{\min\{b,l\}}
        \sum_{s'=0}^{\min\{c,k\}}
        \sum_{u'=0}^{\min\{d,l\}}
        \nonumber\\
        &\times &
        N_{A,B,C,D}^{a,b,c,d,s,u,s',u'}
        \rho_{a,b;c,d}\,\, \rho_{A+k-a,B+l-b,C+k-c,D+l-d},\nonumber
\end{eqnarray}
where
\begin{eqnarray}
        && N_{A,B,C,D}^{a,b,c,d,s,u,s',u'}    =
        2^{-(A+B+C+D+2k+2l)/2}\nonumber\\
        &\times &(-1)^{ A-a+B-b+C-c+D-d+  
        s+u+s'+u'
        } (1-\eta)^{k+l}\nonumber \\
        &\times &\left[
        \binom{a}{s}
        \binom{k}{s}
        \binom{A+k-a}{k-s}
        \binom{A}{a-s}\right]^{1/2} \nonumber\\
&\times &\left[
        \binom{b}{u}
        \binom{l}{u}
        \binom{B+l-b}{l-u}
        \binom{B}{b-u}\right]^{1/2} \nonumber\\
        &\times &\left[
        \binom{c}{s'}
        \binom{k}{s'}
        \binom{C+k-c}{k-s'}
        \binom{C}{c-s'}\right]^{1/2} \nonumber\\
        &\times &\left[
        \binom{d}{u'}
        \binom{l}{u'}
        \binom{D+l-d}{l-u'}
        \binom{D}{d-u'}\right]^{1/2} . \nonumber  \\
\end{eqnarray}
The case $\eta=1$ describes a 
detector with unit efficiency, and this expression reduces
to the one described above.

\subsection{Examples}

The issue here 
is to what extent the noise introduced by imperfect
photon detectors is harmful to the functioning of the protocol. 
Very low detection efficiencies obviously 
render the protocol less effective.
Then, in each step of the iteration more entanglement is lost than 
concentrated. In turns out, however, that it is not necessary to have
detectors with efficiencies very close to unity. Instead, with
efficiencies that are  achievable with present avalanche photodiodes, 
the degree of entanglement can in fact be increased in the actual 
Gaussification step. Astonishingly, for a small number
of iterations of the protocol, the efficiency does not have to
be particularly high at all, as the subsequent cases will exemplify.\\

\begin{figure}[]
\centerline{
\includegraphics[width=7.5cm]{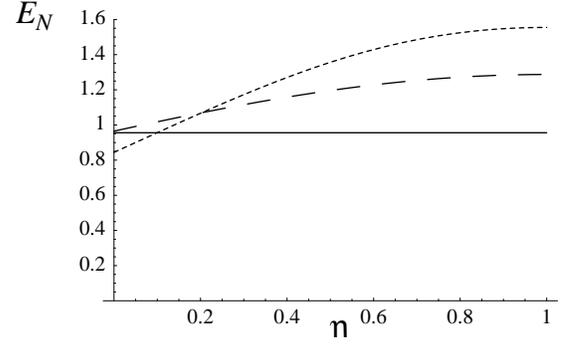}
       }

\vspace{.2cm}
\caption{\label{fig:entanglementnoisy} The logarithmic negativity as a function
of $\eta\in[0,1]$ 
after one and after ten steps of the procedure, for the
initial state as in Example 1 with $\varepsilon=0.7$. The
dashed line is the logarithmic negativity after a single step of
the procedure, the dotted line corresponds to two steps.
The solid line represents the
logarithmic negativity of the initial state prior to
the implementation of the procedure.
}
\end{figure}

\begin{figure}[]
\centerline{
\includegraphics[width=7.5cm]{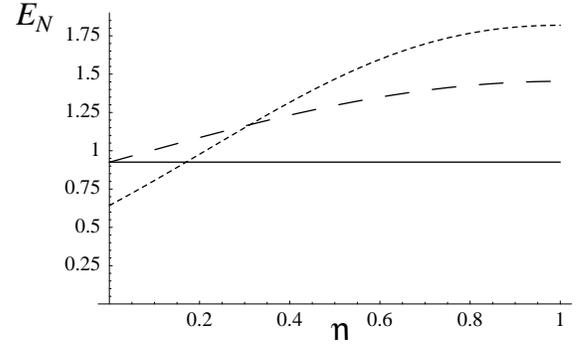}
       }

\vspace{.2cm}
\caption{\label{fig:entanglementnoisy3} As Fig.\
\ref{fig:entanglementnoisy}, but for the initial state
as in Example 1 for $\varepsilon=0.95$.}
\end{figure}

{\it Example 5:} To demonstrate the performance of the procedure with
detectors of non-unit efficiency, we have  computed the logarithmic negativity
for the initial state of Example 1 for $\varepsilon =0.7$
after one step and two steps of
the iteration, as a function of $\eta\in [0,1]$, see Fig.\ 
\ref{fig:entanglementnoisy}. Fig.\ \ref{fig:entanglementnoisy3}
depicts the logarithmic negativity for 
$\varepsilon=0.95$.
The value for the initial state is related to the choice for
the reflectivities and the degree of two-mode squeezing
in the preparatory step. 
Note that for two steps of the procedure
and $80\%$ detection efficiency, the degree of entanglement is
almost doubled.

It becomes clear from these figures that
in these instances, even fairly low
detection efficiencies render the protocol
very effective. Surprisingly, 
for a single step, any detection efficiency
is accompanied with an increase of entanglement, even
for very low ones. This is due to the fact that already without
measurement, no entanglement is lost, due to the particular
structure of the initial state. Whenever the detection efficiency
is larger than zero, the distillation protocol does work.
For a larger number of steps, more and more increase of 
entanglement is to be expected for high detection efficiencies,
while the procedure becomes more and more sensitive with
respect to imperfect detectors. Nevertheless, 
the scheme seems astonishingly robust with respect to imperfections in 
the detectors for a small number of steps. In fact, realistic detection 
efficiencies available with present technology are sufficient
for the functioning of the scheme.

Note that the biggest increase in entanglement per step
will follow the preparatory step as described in Section 
\ref{PreparatoryStep}.
The Gaussification steps in turn will let the state become
Gaussian to arbitrary approximation. The capabilities to serve as
a quantum privacy amplification scheme even in the case of noisy apparata
\cite{HansHans} 
resulting from non-unit detection efficiencies will be discussed in a 
forthcoming publication.\\

\section{Summary and Discussion}\label{Summary}

In this paper we have presented a scheme for distilling
mixed-state continuous-variable entanglement that only
makes use of operations that are accessible in optical systems.
This scheme, the Gaussification scheme applied to mixed states,
is an iterative procedure, in each step of which
modes are brought to interference and are measured using
photon detectors that distinguish between the presence and
the absence of a photon. There is  no essential
need for storing the light in fibers or for mapping the state of
light to atomic degrees of freedom.
It is after all worth noting that the same scheme, when applied
to single-mode systems instead of a two-mode system, would
serve as a procedure to prepare single-mode 
squeezed states of light. The squeezing
would then be entirely due to the measurements performed
in the course of the procedure.

This distillation scheme is on the one hand meant to be a quantum
optical scheme that is close to what can be achieved with present
technology. On the other hand, it is interesting in its own right
from the perspective of the theory of quantum entanglement. It
shows that entanglement concentration is possible from Gaussian to
arbitrarily perfect Gaussian states, yet, to achieve this goal, a
single non-Gaussian preparatory step is necessary. The rest of the
procedure can rely on Gaussian operations, which nevertheless
change the second moments.

We have presented in this paper in detail formal statements
concerning the weak convergence to Gaussian two-mode states. With
several examples we have illustrated the properties of the
procedure, and we have discussed the degree of entanglement and
squeezing that are to be expected. The understanding of the
asymptotic behavior is important from the entanglement theory
points of view. In a practical realisation, in turn, one would be
interested in a powerful scheme where a few steps or a single
step already achieve
the desired goal. 
A crucial limitation in a practical quantum
optical implementation is the non-unit detector efficiency. We
have discussed the constraints in achievable entanglement and
squeezing that are inherited by these limitations. Other
limitations in a scheme consisting of a preparatory step and a
single step of the Gaussification scheme -- and the performance of
protocols with intermediate quantum repeater devices -- will be
investigated in all detail in forthcoming work. 

In the proposed protocol, neither controlled-not operations nor 
photon counters where different 
photon numbers correspond to different
classical signals are required. The technological challenges that 
have to be overcome in the
present proposal are inequivalent to the ones encountered in
the known schemes in the finite-dimensional setting. After all,
it is the hope that the present paper can contribute in a significant
manner to the debate of how to distribute entanglement
over arbitrary distances in the presence of noise.

\section{acknowledgements}

We would like to thank C.H.\ Bennett, J.I.\ Cirac, A.\ Doherty,
L.-M.\ Duan, N.\ Gisin, H.J.\ Kimble, N.\ Korolkova, G.\ Leuchs, W.J.\
Munro, P.-K.\ Lam, E.\ Polzik, and J.\ Preskill for fruitful
discussions on the subject of the paper. Special thanks also to
Ch.\ Silberhorn and I.\ Walmsley for intense and very illuminating
discussions on the experimental feasibility of such a scheme. One
of us (JE) would like to thank J. Preskill and his group for the
kind hospitality at the Institute for Quantum Information at
CalTech, where a significant part of this work has been done. 
This work has been supported by the European Commission 
(EQUIP, QUPRODIS, QUIPROCONE), 
the Alexander-von-Humboldt Stiftung (Feodor-Lynen Grants of SS and
JE), the Deutsche Forschungsgemeinschaft 
DFG, the Engineering and Physical Sciences Research Council EPSRC, 
and Hewlett Packard (CASE award studentship for DEB), and the 
European Science Foundation
programme "Quantum Information Theory and Quantum Computation".
\appendix

\section{Proof of Proposition 1}\label{appa}

This follows immediately
from the fact that the beam splitters are reflected
by the $S\in Sp(8,{\mathbbm{R}})$ given by
\begin{equation}
        S=\frac{1}{\sqrt{2}}\left(
        \begin{array}{cc}
        {\mathbbm{1}}_4 & {\mathbbm{1}}_4 \\
        {\mathbbm{1}}_4  & - {\mathbbm{1}}_4 \\
        \end{array}\right).
\end{equation}
This is the representation of the beam splitter with the
same phase convention as above.
If the first moments $d$ are initially zero, they are also vanishing in later
steps, as the vector of resulting first moments after the first step is given by
  $  d'=(S\oplus S)d=0$.
Let $\gamma$ be the covariance matrix of $\rho$, then
\begin{equation}
        S^{T}(\gamma \oplus \gamma) S=\gamma \oplus \gamma.
\end{equation}
Hence, ${\cal E}(\rho\otimes \rho)=\rho$ holds.
\proofend

\section{Proof of Proposition 2}\label{appb}

The statement that has to be proven is
that except from centered Gaussians no
other two-mode states $\rho$ exist
with the property that $\rho=
c\;{\cal E}(\rho\otimes \rho)$.
Before we will proceed with the proof, we set the notation.
Let us from now on denote the
set of trace-class operators
on ${\cal H}\otimes {\cal H}$ with ${\cal T}$, where ${\cal H} =
{\cal L}^{2}({\mathbbm{ R}})$. The entire
state space
will be denoted by ${\cal S}$. The set of
trace-class operators $\sigma$ with
$\langle 0,0| \sigma | 0,0\rangle= \sigma_{0,0,0,0}=1$ will
be referred to as ${\cal N}$.
Furthermore, we introduce the
set of centered two-mode Gaussians ${\cal G}\subset {\cal T}$
by
\begin{eqnarray}
{\cal G}= \left\{
\rho: \rho=\int d^{4}\xi e^{-\xi^{T}\Sigma^T
A \Sigma
\xi/4} W_{-\xi},\,
\text{$A\geq 0$, \text{ real}  }\right\},\nonumber\\
\end{eqnarray}
which includes also the two-mode
Gaussians with vanishing first moments with
sub-Heisenberg variance.
These Gaussians with sub-Heisenberg variance,
meaning that $A+ i \Sigma$ is not a positive matrix,
correspond to elements of ${\cal T}$ which are not positive
and hence not states.
The task is to show that for each $\rho\in{\cal S}$
for which there exists a $c>0$ such that
\begin{equation}\label{fpn}
    \rho=c {{\cal E}}(\rho\otimes \rho)
\end{equation}
it follows that
$\rho\in {\cal G}$.

The first step is to consider  $\rho\in {\cal T}$
that are the solutions of Eq.\ (\ref{fpn})
with some $c>0$ in the number basis.
For simplicity, let us fix normalization
by considering
$\sigma= \rho/ \rho_{0,0,0,0}=\rho/ \langle 0,0|
\rho|0,0\rangle$,
which are the solutions of $\sigma ={\cal E}(\sigma\otimes \sigma)$.
The relevant observation is that for any
$i,j,k,l\in {\mathbbm{N}}_0$, the number
\begin{equation}
    \sigma_{i,j,k,l}=\langle i,j|\sigma|k,l\rangle
\end{equation}
is a polynomial of degree
\begin{equation}
        D\leq \max\{i,j,k,l\}
\end{equation}
in the complex numbers
\begin{eqnarray}
    &&\sigma_{1,0,1,0},
    \sigma_{0,1,0,1},
    \sigma_{1,0,0,1},\\
    &&\sigma_{2,0,0,0},
    \sigma_{0,2,0,0},
    \sigma_{1,1,0,0}.
\end{eqnarray}
This can be proven by induction, on using the
above map Eq.\ (\ref{r2})
in terms of the number basis.
For $\sigma_{i,j,k,l}$ with
$i,j,k,l\leq 2$ this can be seen by direct inspection.
Then, assume that the above statement is true for all $\sigma_{i,j,k,l}$
with $i,j,k,l\leq m$ for some $m>2$. One can then consider first
$\sigma_{m+1,0,0,0}$, to immediately find that
\begin{eqnarray}
        \sigma_{m+1,0,0,0} &=& \sigma_{m+1,0,0,0} \nonumber\\
        &\times &
        2^{-(m+1)/2} \left(
        (-1)^{m+1} +1
        \right)+ r,
\end{eqnarray}
where $r$ is a polynomial in the entries of $B$.
Similarly, one can proceed
by showing that the statement is true for
all
$\sigma_{i,j,k,l}$
with $i,j,k,l\leq m+1$, which
is the induction step \cite{Zeroremark}. This means
that the set of $\rho\in{\cal T}$ that satisfy the above
fixed-point condition Eq.\ (\ref{fpn}) gives rise to
a ten-dimensional
manifold -- just as the set of centered Gaussians. The remainder
of the proof is merely concerned with showing that indeed,
any such fixed point is nothing but a centered Gaussian.

Equivalently, $\sigma_{i,j,k,l}$ is a polynomial of
degree $D$ in the entries of a real symmetric matrix
$4\times 4$-matrix $B$, by setting
\begin{eqnarray}
        \sigma_{1,0,1,0}&=&1-B_{1,1}-B_{2,2}, \nonumber\\
        \sigma_{0,1,0,1}&=& 1-B_{3,3}- B_{4,4}, \nonumber\\
        \sigma_{1,0,0,1}&=& - B_{1,3} - B_{2,4} + i(B_{1,4}- B_{2,3}), \nonumber\\
        \sigma_{2,0,0,0}&=& 2^{-1/2} (B_{1,1} -B_{2,2}+2 i B_{1,2}), \nonumber\\
        \sigma_{0,2,0,0}&=& 2^{-1/2} (B_{3,3} - B_{4,4} + 2 i B_{3,4}), \nonumber\\
        \sigma_{1,1,0,0}&=& B_{1,3} - B_{2,4}+
        i(B_{1,4}+B_{2,3}).\label{en}
\end{eqnarray}
This particular choice will be motivated below.
As the affine map (\ref{en}) is invertible,
each  $\sigma \in {\cal N}$ that is the
solution of Eq.\ (\ref{fpn})
is up to normalization
uniquely
characterized by the entries of a real
symmetric $4\times 4$-matrix $B$. This gives rise
to a map
\begin{equation}
        F:{\cal N}\longrightarrow M_{4\times 4}
\end{equation}
where $M_{4\times 4}$
denotes the set of real symmetric $4\times 4$-matrices.
$F$ is a bijection relating ${\cal N}$ and $F({\cal N})\subset M_{4\times 4}$.

Let $C_{4\times 4}\subset M_{4\times 4}$
be the set of
real symmetric $4\times 4$-matrices $B$
for which
$\sigma = F^{-1}(B)$
is moreover positive, $\sigma\geq 0$.
As any centered Gaussian
\begin{equation}\label{CGauss}
\rho=\int d^{4}\xi e^{-\xi^{T}\Sigma^T
A \Sigma
\xi/4} W_{-\xi}
\end{equation}
with $A\geq 0$
satisfies $\rho={\cal E}(\rho\otimes \rho)$,
there clearly
exists a matrix $B\in C_{4\times 4}$ with
\begin{equation}
        \rho/\rho_{0,0,0,0}=F^{-1}(B).
\end{equation}
This matrix $B$ can be identified as
being given by
\begin{equation}\label{connec}
        B= (A +{\mathbbm{1}}_{4})^{-1}.
\end{equation}
This can be seen as follows:
The number
$\rho_{0,0,0,0}$
can be expressed as
\begin{eqnarray}
        \rho_{0,0,0,0} &=&
        \frac{1}{\pi^2}
        \int d^4 \xi \exp(-\xi^T \Sigma^{T} A
        \Sigma \xi/4) \langle 0,0 | W_{-\xi} | 0,0\rangle\nonumber\\
        &=& \frac{1}{\pi^2} \int
        d^4 \xi \exp(-\xi^T \Sigma^{T}  (A
        +{\mathbbm{1}}_{4})\Sigma \xi /4)\nonumber\\
        &=& (2/| A+{\mathbbm{1}}_4|)^{1/2}= |2 B|^{1/2},
\end{eqnarray}
where $|.|$ denotes the determinant.
Setting again $\sigma=\rho/\rho_{0,0,0,0}$ for $\rho$ as in Eq.\ (\ref{CGauss}),
on using the CCR for Weyl operators, one arrives after a
few steps at
\begin{eqnarray}
        \sigma_{0,1,0,1} &=& 1-
        \frac{1}{\pi^2}
        \int d^4 \xi \exp(-\xi^T \Sigma^{T}  A
        \Sigma \xi/4)\nonumber\\
        &\times & (\xi_3^2 + \xi_4^2)/2,
\end{eqnarray}
that is, $\sigma_{0,1,0,1}=1- ( B_{3,3} + B_{4,4})/2$.
In the same way
one finds that Eqs.\ (\ref{en}) hold,
and hence at
$F^{-1}(B)=\rho/\rho_{0,0,0,0}$.

What remains to be shown is that for any $B\in C_{4\times 4}$
there exists an $A\geq 0$ with $B=(A+{\mathbbm{1}}_{4})^{-1}$.
Such an $A$ can always be found if it is true that
for any $B\in C_{4\times 4}$
\begin{equation}
    B\leq {\mathbbm{1}}_{4},\,\,\,\,
    |B|\neq 0.
\end{equation}
This is indeed the case:
if $|B|=0$, then $\langle 0,0 |F^{-1}(B)|0,0\rangle =0$, and
hence,
$F^{-1}(B)\neq {\cal N}$.
If $B\leq {\mathbbm{1}}_{4}$,
we can without loss of generality assume that $B_{1,1}>1$
(which is always achievable with an appropriate basis change).
Then $\langle 1,0|
F^{-1}(B)|1,0\rangle<0$, and hence, $F^{-1}(B)\neq {\cal N}$.
So we arrive at the conclusion that any (unnormalized
state) $F^{-1}(B)\in{N}$ with $B\in C_{4\times 4}$
can be realised as
a centred Gaussian. Hence, we can conclude that
there are no non-Gaussian
fixed points, which was the statement that had to be shown.
\proofend

\section{Proof of Proposition 3}\label{appc}

Let us assume that $\langle 0,0|\rho |0,0\rangle=
\rho_{0,0,0,0}=1$. The first observation is that
\begin{eqnarray}
    \rho^{(i)}_{1,0,1,0}&= \rho^{(1)}_{1,0,1,0},\,\,
 \rho^{(i)}_{0,1,0,1}&= \rho^{(1)}_{0,1,0,1},\\
  \rho^{(i)}_{1,0,0,1}&= \rho^{(1)}_{1,0,0,1},\,\,
   \rho^{(i)}_{2,0,0,0}&= \rho^{(1)}_{2,0,0,0},\\
    \rho^{(i)}_{0,2,0,0}&= \rho^{(1)}_{0,2,0,0},\,\,
     \rho^{(i)}_{1,1,0,0}&= \rho^{(1)}_{1,1,0,0},
\end{eqnarray}
for all $i\geq 1$. In other words, these numbers change
only in the first step, but not in further steps.
This
is an immediate consequence of the iteration map
${\cal E}$ as in Eq.\ (\ref{nl1}). The coefficients
$\rho^{(1)}_{1,0,1,0}$, $\rho^{(1)}_{0,1,0,1}$, $\rho^{(1)}_{1,0,0,1}$,
$\rho^{(1)}_{2,0,0,0}$, $ \rho^{(1)}_{0,2,0,0}$, $\rho^{(1)}_{1,1,0,0}$
specify the resulting second moments of  the state
according to $F(\rho^{(1)})$. The matrix of the affine map in Eq.\ (\ref{superlong})
is the inverse of the map Eq.\ (\ref{en}). By induction, it can be shown that each
$\rho^{(i)}_{a,b,c,d}$ for $a,b,c,d\in{\mathbbm{N}}_0$
converges pointwise to $\langle a,b| F(\rho^{(1)})|c,d\rangle$ as $i\rightarrow \infty$.
$F(\rho^{(1)})$ is a positive trace-class operator, if and only if the second moments
of the resulting Gaussian satisfy the Heisenberg uncertainty relation.  \proofend

\section{Proof of Proposition 4}\label{appd}

The starting point is the observation that any two-mode
pure Gaussian state
\begin{equation}
        |\psi\rangle\langle\psi|= \sum_{s,t,n,m=0}^\infty \omega_{s,t;n,m}
        |s,t\rangle\langle n,m|
\end{equation}
with vanishing first moments
has the property that
$       \omega_{1,0,1,0}=0$. One way to see this is as follows:
On the level of covariance matrices,
and pure two-mode covariance matrix can be decomposed into
\begin{equation}
    \gamma = O D O^{T},
\end{equation}
where $D=\text{diag}(d_{1},1/d_{1},d_{2},1/d_{2})$,
$d_1,d_2\in{\mathbbm{R}}^+$,
and
$O\in Sp(4,{\mathbbm{R}})\cap SO(4)$. Written in terms of
state vectors one obtains
\begin{equation}
    |\psi\rangle = U \left(S(d_{1})\otimes S(d_{2})\right) |0,0\rangle,
\end{equation}
where $U$ is the unitary representing the passive transformation $O$,
and $S(d_{1})$ and $S(d_{2})$ denote
the single-mode squeezing operators. Representing this in terms of
creation and annihilation operators, one arrives at the above
property.
%
Hence, the above property enforces that
\begin{eqnarray}\label{conve}
        \rho^{(i)}_{1,0,1,0}\longrightarrow 0\, \text{ as }\, i\rightarrow \infty
\end{eqnarray}
must necessarily hold. 
Let us again consider $\sigma=\rho/\rho_{0,0,0,0}$, and
\begin{equation}
        \sigma'=
	{\cal E}(\sigma\otimes \sigma)= \sum_{s,t,n,m=0}^\infty
	\sigma'_{s,t;n,m}
        |s,t\rangle\langle n,m|
\end{equation}
be the positive operator after the first step. Then
\begin{equation}
        \sigma'_{1,0,1,0}=  
	\sigma_{1,0,1,0} - \sigma_{1,0,0,0}
	\sigma_{0,0,1,0}.
\end{equation}
We have that
$\sigma'_{1,0,0,0}=0=\sigma_{0,0,1,0}^{(1)}$ for all initial
two-mode tracece-class operators $\sigma$ with $\sigma_{0,0,0,0}=1$,
as follows from the iteration
map. Therefore, it is clear
that Eq.\ (\ref{conve}) can only be true if already
\begin{equation}\label{q1}
        \sigma_{1,0,1,0} = \sigma_{1,0,0,0} \sigma_{0,0,1,0} =
	|\sigma_{1,0,0,0}|^2
\end{equation}
holds. This is the first requirement that emerges if one requires
exact convergence to a pure Gaussian state.
In just the same manner one can first conclude that
\begin{equation}
    \lim_{i\rightarrow\infty} \sigma^{(i)}_{0,1,0,1}=0,\,\,
    \lim_{i\rightarrow\infty} \sigma^{(i)}_{1,0,0,1}=0
\end{equation}
must hold, and therefore,
\begin{eqnarray}\label{q2}
        \sigma_{0,1,0,1}=|\sigma_{0,1,0,0}|^2,\,\,
        \sigma_{1,0,0,1}=|\sigma_{1,0,0,0}|^2.
\end{eqnarray}
In order to proceed, let us take
$\sigma'= \rho'/\rho'_{0,0,0,0}$
as input and starting point. It has the property that
\begin{eqnarray}\label{extrac}
        \sigma'_{1,0,1,0}&=& \sigma'_{0,1,0,1}=
        \sigma'_{1,0,0,1}=0,\\
        \sigma'_{0,0,0,0}&=& 1.
\end{eqnarray}
According to Proposition 3,
to which final state the procedure converges to
is only dependent on the numbers
$\sigma'_{1,1,0,0}$, $\sigma'_{2,0,0,0}$, and
$\sigma'_{0,2,0,0}$.
%
Hence,
the question whether  $\rho^{(i)}$
satisfying Eq.\ (\ref{extrac}) converges
to an element in ${\cal S}$ as $i\rightarrow \infty$
is equivalent to asking whether
\begin{equation}
{\cal E}^{(i)}(|\phi\rangle\langle\phi|)/
\text{tr}[ {\cal E}^{(i)}(|\phi\rangle\langle\phi|)]
\end{equation}
with state vector
\begin{eqnarray}
    |\phi\rangle = \beta_{2,0} |2,0\rangle + \beta_{0,2}|0,2\rangle+
    \beta_{1,1} |1,1\rangle+ |0,0\rangle
\end{eqnarray}
converges to an element in ${\cal S}$.
This question, in turn,
has been answered in Ref.\  \cite{Gaussify}. On using the results
of Proposition 4 in Ref.\ \cite{Gaussify}, one arrives at necessary
and sufficient conditions for convergence of unnormalized
states satisfying Eq.\ (\ref{extrac}) is
\begin{equation}\label{jc}
    \left\|\left(
    \begin{array}{cc}
        \sqrt{2} \sigma'_{2,0,0,0} & \sigma'_{1,1,0,0}\\
        \sigma'_{1,1,0,0} & \sqrt{2} \sigma'_{0,2,0,0}
    \end{array}\right)\right\|_{\infty}<1,
\end{equation}
where $\|.\|_{\infty}$ denotes the spectral norm.
To see what Eq.\
(\ref{jc}) reflects in the first step, one has to apply the iteration
map, giving rise to the condition for the
initial state $\rho$,
\begin{eqnarray}\label{secc}
    &&\left\|\left(
    \begin{array}{cc}
      \sqrt{2} \frac{\rho_{2,0,0,0}}{\rho_{0,0,0,0}}
       - \frac{\rho_{1,0,0,0}^{2}}{\rho_{0,0,0,0}^{2}} &
       \frac{\rho_{1,1,0,0}}{\rho_{0,0,0,0}}
       -\frac{\rho_{1,0,0,0}\rho_{0,1,0,0}}{\rho_{0,0,0,0}}\\
       \frac{\rho_{1,1,0,0}}{\rho_{0,0,0,0}}
       -\frac{\rho_{1,0,0,0}\rho_{0,1,0,0}}{\rho_{0,0,0,0}}
       &
       \sqrt{2} \frac{\rho_{0,2,0,0}}{\rho_{0,0,0,0}} -
       \frac{\rho_{1,0,0,0}^{2}}{\rho_{0,0,0,0}^{2}}
    \end{array}\right)\right\|_{\infty}<1.\nonumber\\
\end{eqnarray}
Combining Eq.\ (\ref{secc}) with Eqs.\ (\ref{q1}),
(\ref{q2})
yields the above necessary and sufficient conditions for
convergence to a pure centred two-mode Gaussian.
Note that the convergence is meant in the sense of
weak convergence, and not in the norm sense.
\proofend


\begin{thebibliography}{99}
\bibitem{QuantNote}
        It should be mentioned
        that the quantification of entanglement for infinite-dimensional
        quantum systems is a subtle issue \cite{QuantCont,QuantForm,Infi}.
        The minimization necessary to evaluate most measures
        of entanglement is typically not feasible, and without further assumptions
        they are not even trace-norm continuous \cite{QuantCont}. For
        those two-mode Gaussian
        states the sequence of states converges to in the weak sense, the entanglement
        of formation in the infinite setting \cite{QuantCont} is available in principle: it
        has been shown in Ref.\ \cite{QuantForm} that for the symmetric Gaussians
        that we encounter here, the entanglement of formation is identical to the
        Gaussian entanglement
        of formation. For the intermediate steps this clearly
        does not hold, however.
        In this paper, we take a pragmatic approach and choose
        the computable logarithmic negativity as the appropriate functional
        that quantifies the degree of entanglement.
\bibitem{Zeroremark}
        Note that in fact
        $\sigma_{i,j,k,l}=0$
        if $i+j+k+l$ is an odd number,
        as can again be shown by induction.
\end{thebibliography}
\end{document}